\documentclass[12pt,cites,graphicx]{article}
\usepackage{epsfig}
\usepackage{color}

\title{ Computational Assessment of the Entropy of Solvation of 
        Small-Sized Hydrophobic Entities }

\author{ Reema Mahajan,$^a$ 
         Dieter Kranzlm\"uller,$^b$ 
         Jens Volkert,$^b$ \\
         Ulrich H. E. Hansmann$^{c,d}$ 
         and Siegfried H\"ofinger$^{c}$\\[3mm]
         $^a$ Indian Institute of Technology, Delhi, \\ 
              Department of Chemical Engineering, \\
              Hauz Khas, New Delhi-16, India. \\
              E-mail: reema.mahajan@gmail.com\\[1mm]
         $^b$ Johannes Kepler University of Linz  \\,
              Institute of Graphics and Parallel Processing, \\
              Altenberger Stra\ss e 69, A-4040, Linz, Austria. \\
              E-mail: \{kranzlmueller, volkert\}@gup.jku.at\\[1mm]
         $^c$ Michigan Technological University, \\
              Department of Physics, \\
              1400 Townsend Drive, Houghton, MI, 49331-1295,USA. \\
              E-mail: \{hansmann, shoefing\}@mtu.edu\\[1mm]
         $^d$ John von Neumann Institute for Computing, FZ J\"ulich, \\
              52425 J\"ulich, Germany. \\
       }

\begin{document}
\maketitle
\renewcommand{\thefootnote}{\fnsymbol{footnote}}

\noindent 
A high level polarizable force field is used to study the temperature
dependence of hydrophobic hydration of small-sized molecules from 
computer simulations. Molecular dynamics simulations of liquid water 
at various temperatures form the basis of free energy perturbation 
calculations that consider the onset and growth of a repulsive sphere. 
This repulsive sphere acts as a model construct for the hydrophobic 
species. In the present study an extension is pursued to all in all 
seven independent target temperatures starting close to the freezing 
point and ranging up to almost the boiling point of liquid water at 
standard conditions. Care is taken to maintain proper physico-chemical
model description by cross-checking to experimental water densities 
at the selected target temperatures. The polarizable force field 
description of molecular water turns out to be suitable throughout 
the entire temperature domain considered. Derivatives of the computed 
free energies of hydrophobic hydration with respect to the temperature 
give access to the changes in entropy. In practice the entropy 
differential is determined from the negative of the slope of tangential 
lines formed at a certain target temperature in the free energy 
profile. The obtained changes in entropy are of negative sign for small 
sized cavities, hence reconfirm basic ideas of the {\it Lum 
Chandler Weeks} theory on hydrophobic hydration of small-sized solutes.

\section{Introduction}
The hydrophobic effect is widely believed to play a decisive role
in protein folding, one of the key challenges in biophysical 
science and research of today \cite{chandler,liu,despa}. Theoretical 
studies on the hydrophobic effect are of great relevance to a broad 
range of biosciences and a deeper understanding of hydrophobicity 
could certainly have its beneficial influence on many central questions 
in current biophysical research. Among others, {\it Lum Chandler Weeks} 
(LCW) theory of hydrophobicity \cite{lum} has received widespread 
appreciation. LCW theory describes the hydrophobic effect in terms 
of reorganizational work due to maintainance of a hydrogen bond network 
established between individual water molecules. A difference is made 
between small-sized hydrophobic solutes (volume dependence) and large-sized 
hydrophobic solutes (surface area dependence). Water molecules are believed 
to re-arrange appropriately around small-sized hydrophobic solutes thereby
inducing the formation of clathrate-like substructures without destruction 
of the hydrogen bond network. In contrast, large-sized hydrophobic solutes 
are thought to enforce a complete re-arrangement of the hydrogen bond 
network adjacent to the hydrophobic solute, which in turn leads to 
agglomeration, aggregation and precipitation of the large-sized hydrophobic 
molecules. Such a picture of hydrophobicity would render the change in 
entropy for the process of solvating small-sized hydrophobic solutes to be 
negative in sign (increase in order), and the opposite for large-sized 
hydrophobic molecules (decrease in order). Hence, two immediate questions  
arise naturally: i) can computer simulations verify the claim of entropic 
drop for hydration of small-sized hydrophobic solutes ?  ii) are current 
model descriptions of molecular water able to reproduce the physics of 
hydrophobicity correctly \nolinebreak ? Although complementary, these two 
questions have to be addressed by atomistic computer simulations 
of hydrophobic solutes in aqueous solution.

Computer simulations have become a valuable tool in the study of 
hydrophobicity \cite{pratt,hummer,floris,choudhury,ashbaugh}. Among other
techniques, free energy calculations have been introduced and advanced to 
directly study $\Delta$G trends for various physico-chemical processes 
\cite{zwanzig,postma,simonson}. One such possible process is to investigate 
the $\Delta$G corresponding to the introduction, onset and growth of a 
repulsive sphere located in the center of a simulation cell filled with 
water molecules. In such a model, the repulsive sphere stands as a 
representative of an artificial hydrophobic solute. The associated free 
energy change is known as the cavitation free energy, 
$\Delta$G$^{cav}$ \cite{tomasi}. Postma and coworkers have introduced 
this type of calculation as one of the early examples of {\it Free Energy 
Perturbation} calculations (FEP). Their approach is known as the 
{\it Overlapping Spheres Technique} (OST) \cite{postma}.  Recent 
re-evaluations and variations of the OST in the context of hydrophobicity 
have been summarized in \cite{hoefi}. The advantage of true estimates of 
$\Delta$G becomes evident when looking at the derivative with respect to 
temperature, 
$\frac{\partial} {\partial T} (\Delta G) = \frac{\partial}{\partial T} 
( \Delta H - T \Delta S) = - \Delta S $, 
i.e. from a record of different $\Delta G$ values at different temperatures 
the change in entropy can  be determined from the negative slope of the 
tangent formed at a particular point. Thus if the LCW arguments hold true, 
then for solvating small-sized hydrophobic solutes one would obtain a 
bell-shaped curve of the temperature plot of $\Delta$G, for otherwise the 
slope of the tangent at room temperature can not become positive, 
hence $\Delta$S not negative. 
\textcolor{red}{However, high temperature simulations in the molecular
mechanics description need to be carefully cross-checked against 
experimental data whether they still can provide reasonable representations 
of the native state. Therefore, only if one can confirm that a set of high 
temperature simulations still takes place at reasonable physico-chemical 
conditions, an estimate of entropies can be obtained. For the type of 
atomistic interaction governing hydrophobic phenomena, a proper account of 
the liquid water density at high temperatures is the most essential 
precondition for pursuing estimates of the entropy \cite{paschek,paschek1}.}

Model descriptions of water have a great influence on the outcome of
biomolecular simulations \cite{zhou,pande} but the employment of prominent 
water models \cite{jorgensen,berendsen} is common practice in present 
biophysical research. A specific high level description of molecular 
water has been proposed with the use of polarizable models \cite{ren}. 
The just cited AMOEBA model has been shown to exhibit excellent 
description of the temperature and pressure dependence of water 
\cite{ren1}. AMOEBA water was successfully applied in describing 
aspects of the hydrophobic effect \cite{hoefi1}. Since this 
previous study was focussing on room temperature behavior, an extension 
towards high temperature repeats is straightforward as it could lead to 
new insight into fundamental principles of hydrophobicity. Other studies 
on the entropy of hydrophobic hydration were mainly based on rigid water 
models so far \cite{ghosh,rajamani}.

The present article reports cavitation free energy calculations performed
at seven individual temperatures in the range of 277 K to 370 K. The 
polarizable AMOEBA water model \cite{ren} is used and Ewald summation is 
applied within the {\it Molecular Dynamics} simulations (MD) that form the 
basis of the FEP calculation of the OST approach. Emphasis is placed on 
verification of proper physico-chemical model description at high 
temperatures by comparison to experimentally obtained trends of liquid 
water densities. Entropic changes for this process are derived
from the temperature dependence, that is the entropy of hydrophobic 
hydration is estimated by means of computer simulations. The computational 
demand of this study is on the order of three CPU years on decent 
architectures (Itanium 1.4 GHz) \textcolor{red}{ \cite{benjamin} } and can 
only be satisfied from massive employment of grid computing systems, such 
as for example the Austrian grid \cite{AutGrid}.

\section{ Methods }
\subsection{ Simulation Cell Set Up } 
\label{section_setup}
Seven individual cubic boxes composed of 6 x 6 x 6 grid cells were formed, 
where the sub-volume of the grid cells was adjusted to reproduce the 
experimental liquid water density corresponding to a chosen target 
temperature. Seven target temperatures were selected, 277 K, 300 K, 315 K, 
330 K, 345 K, 365 K and 370 K. The structure of a single water molecule was 
optimized and then periodically translated and copied to each of the centers 
of the grid cells. Thus all simulation cells contained 216 water molecules.
After initial construction the systems were minimized and subjected to
simulated annealing using 2000 steps of 1.0 fs each to approach 1000 K
peak temperature before linearly cooling down to one of the seven target 
temperatures. Volume modification during simulated annealing was less than 
1 \% when compared to the box dimensions upon start-up. All calculations were
performed with the TINKER package for molecular modeling version 4.2 
\cite{ponder}.

\subsection{ MD/FEP Calculations } 
24 individual MD-trajectories per chosen target temperature (see section 
\ref{section_setup}) were recorded. The TINKER package \cite{ponder} was 
used for all computations. Polarizable AMOEBA force field parameters for 
water \cite{ren} were employed. 
\textcolor{red}{ AMOEBA works on the basis of self-consistent induced atomic
dipoles. }
NpT ensembles were selected at 1.0 atm 
target pressure with the temperature/pressure coupling method due to 
Berendsen \cite{berendsen1} which accounts for box-size changes when the 
volume of the repulsive cavity is introduced.  
\textcolor{red}{ Default coupling constants to the thermostat and barostat
were used, that is 0.1 ps and 2.0 ps respectively. }
\textcolor{red}{ Other choices could have been made \cite{andersen}, 
especially with respect to the ``Flying ice cube" phenomenon \cite{cheatham},
but here we tried to follow closely the protocol used during AMOEBA
development. }
Time steps of 1.0 fs were 
employed, no restraints/constraints were applied and Ewald summation was 
used. Two types of perturbation potentials were 
introduced, 
\begin{equation}
   \label{eq1}
   V_{rep} = \lambda \left( \frac{B^*}{r} \right)^{12}
\end{equation}
and 
\begin{equation}
   \label{eq2}
   V_{rep}^{mod} = \frac{\lambda^{12}}
                           {\left[
                                  0.3(1-\lambda)^2 
                                  + \left( \frac{r}{B^{*}=1.0} \right)^{6}
                            \right]^2}
\end{equation}
to avoid discontinuities when the thermal radius of the repulsive cavity, 
$B^{*}$, approaches 0 \AA \, \cite{beutler,simonson1}. Parameter $\lambda$ 
in equations (\ref{eq1}) and (\ref{eq2}) describes the degree of perturbation
and assumes values from $\lambda=0$ (unperturbed) to $\lambda=1$ (fully 
perturbed). All technical details concerning the perturbation potential 
are given in \cite{hoefi2}. Trajectories were recorded for 100 ps each and
the perturbations shown in equations (\ref{eq1}) and (\ref{eq2}) were 
calculated and stored every time step. As reported previously, the 
$\lambda=0.5$ simulation applying $V_{rep}^{mod}$ of equation (\ref{eq2}) 
is quasi-unperturbed and was used to monitor the time evolution of the 
macroscopic liquid water density. Data evaluation was started only after 
stable levels of the liquid water density had been reached (approximately 
after 40 to 50 ps, see Table \ref{table1}). 
\textcolor{red}{ The $\lambda=0.5$ simulation has also been analyzed with 
respect to maintainance and changes in the Hydrogen Bond network, as 
outlined in \cite{wernet,xenides} (see Supplementary Material). }

\subsection{ Data Evaluation } 
For each of the 7 selected target temperatures (see section 
\ref{section_setup}) the recorded 25 trajectories were used for application 
of Zwanzig's \cite{zwanzig} formula
\begin{equation}
  \label{eq3}
  \Delta G(\lambda_i) 
  = 
  -k_B T \hspace{0.3cm} 
  ln 
  \left< 
        e ^{
             -\frac{1}{k_B T}
             \left[
                   {\mathcal H}(\lambda_i+\delta\lambda)
                   -{\mathcal H}(\lambda_i)
             \right]
           }
  \right>_{\lambda_i} 
\end{equation}
with $< \; >$ depicting a thermodynamic average, $k_B$ being the Boltzmann 
constant and ${\mathcal H}$ the total energy of the system. The average is 
formed at $\lambda_i$ with small perturbations $\delta\lambda$ around 
$\lambda_i$. The OST is applied similarly to the description given in 
\cite{hoefi2}. Care has been taken of using proper $k_B T$ factors
for all the seven different temperatures mentioned above. This will effect
equation (\ref{eq3}) and the conversion from repulsive radii, $B$, to thermal
radii, $B^*$, i.e. $B^* = B (1 k_B T)^{-\frac{1}{12}}$. Perturbations 
exceeding energies of $2 k_B T$ were not considered for total averages 
(FEP requirement). There are however always sufficiently enough alternative 
overlap combinations in all the simulations to smoothly connect two adjacent 
repulsive spheres and get statistical averages. Obtained raw data from
the OST were fitted with polynomials of degree 2 and corresponding
coefficients $k_0, k_1$ and $k_2$ are summarized in Table \ref{table1}.
\textcolor{red}{ Resulting cavitation free energies from the coefficients
in Table \ref{table1} are expressed in units of kcal/mol 
(1 kcal = 4.184 kJ). }

\section{ Results }
\subsection{ AMOEBA based computer simulations of molecular liquid water 
             confirm LCW ideas of the unit volume entropic change for 
             solvating small-sized hydrophobic solutes }
\begin{figure}[ht]
\centering
\includegraphics[scale=1.6]{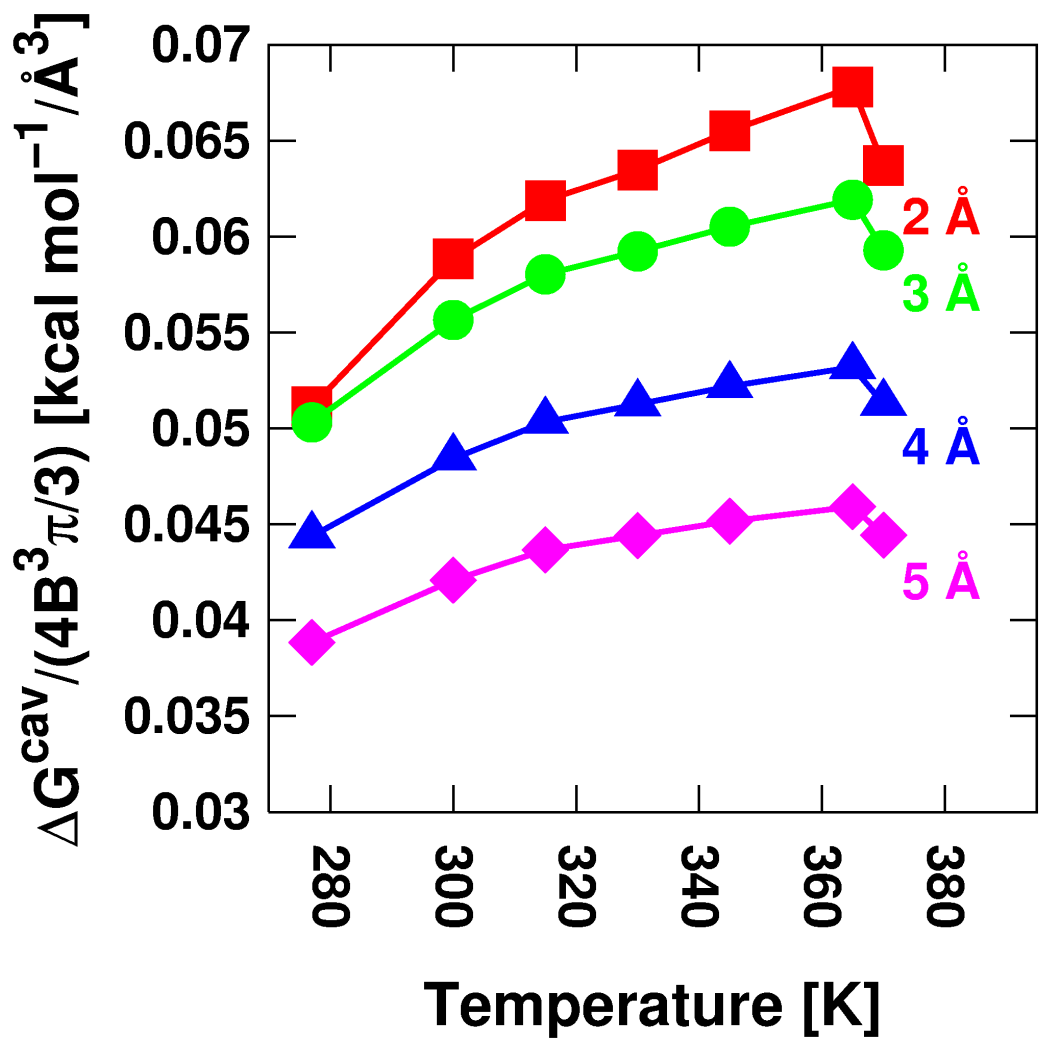}
\caption[  Temperature dependence of the unit volume free
           energy of cavitation for small repulsive spheres of growing
           size. Small-sized cavities of radii B = 2 \AA \, (squares) 
           and B = 3 \AA \, (discs) exhibit room temperature tangents 
           with positive slope that correspond to a unit volume 
           $\Delta$S of negative sign. Similar analysis
           of tangential slopes for larger volumes, i.e. B = 4 \AA \,
           (triangles) and B = 5 \AA \, (diamonds)
           demonstrates that the magnitude of the slopes decreases
           at 300 K when the size of the solute becomes bigger. ]
           { \label{fig1} Temperature dependence of the unit volume free
           energy of cavitation for small repulsive spheres of growing
           size. Small-sized cavities of radii B = 2 \AA \, (squares) 
           and B = 3 \AA \, (discs) exhibit room
           temperature tangents with positive slope that correspond to
           a unit volume $\Delta$S of negative sign. Similar analysis
           of tangential slopes for larger volumes, i.e. B = 4 \AA \,
           (triangles) and B = 5 \AA \, (diamonds)
           demonstrates that the magnitude of the slopes decreases
           at 300 K when the size of the solute becomes bigger. }
\end{figure}

MD/FEP calculations based on the polarizable force field AMOEBA for the 
computation of cavitation free energies in liquid molecular water are 
carried out at seven different target temperatures. The results are
fitted similarly to previous calculations \cite{hoefi2} and resulting
coefficients $k_0$, $k_1$, $k_2$ are summarized in Table \ref{table1} 
({\it revised Pierotti Approach}, rPA). The coefficients are used to 
calculate temperature trends for growing cavities of perfectly spherical 
shape. Figure \ref{fig1} shows a comparison of these data in reduced 
energy units normalized to unit volumes. This type of reduced units allows 
one to immediately compare unit volumes of 1 cubic \AA \, to each other. 
Unit volume cavitation free energies of small-sized cavities (squares 
and discs in Figure \ref{fig1}) clearly exhibit a positive slope for 
the tangent formed at room temperature data points. This indicates a drop 
in entropy for the process of hydrating a hydrophobic volume of 1 cubic 
\AA. Similar constructions of tangents for larger-sized cavities  
(triangles and diamonds in Figure \ref{fig1}) show that the positive 
slopes become smaller as cavities grow. Thus high level force field based 
MD/FEP calculations verify the anticipated change in unit volume entropy 
following LCW theory for small-sized hydrophobic solutes (see for example 
Figure 2 in \cite{huang} for comparison to Figure \ref{fig1} of the present 
work).

\subsection{ Cavitation entropies of small sized hydrophobic solutes
             decrease steadily with increasing cavity size }
\begin{figure}[ht]
\centering
\includegraphics[scale=1.6]{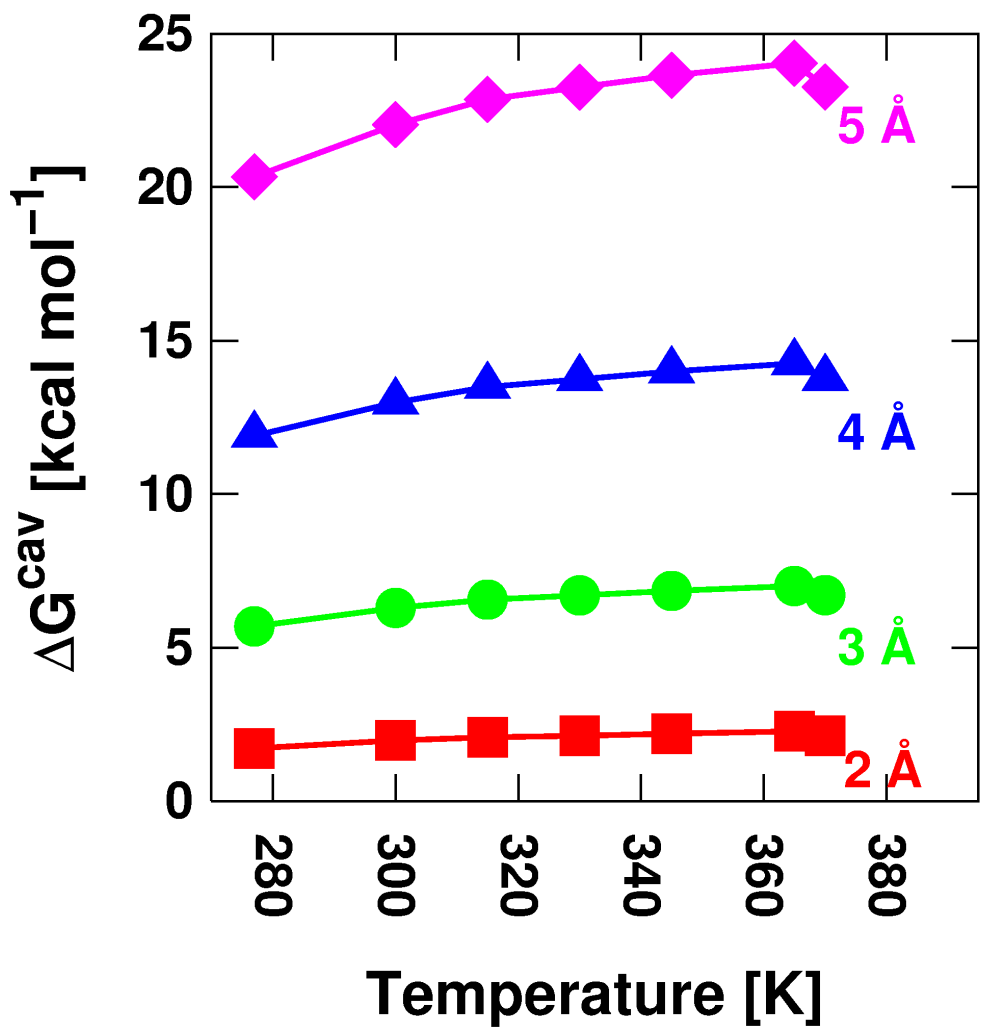}
\caption[  Temperature dependence of the full size free
           energy of cavitation for small repulsive spheres of growing
           size. Room temperature tangents exhibit positive slopes that
           correspond to a $\Delta$S of cavitation of the entire cavity
           volume with negative sign. The magnitude of the slope of
           tangential lines increases with growing cavity size, i.e.
           when following cavities of  radii B = 2 \AA \, (squares),
           B = 3 \AA \, (discs), B = 4 \AA \, (triangles) and
           B = 5 \AA \, (diamonds). Inversion in the sign of the
           slopes of tangential lines is seen close to the boiling point. ]
           { \label{fig2} Temperature dependence of the full size free
           energy of cavitation for small repulsive spheres of growing
           size. Room temperature tangents exhibit positive slopes that
           correspond to a $\Delta$S of cavitation of the entire cavity
           volume with negative sign. The magnitude of the slope of
           tangential lines increases with growing cavity size, i.e.
           when following cavities of  radii B = 2 \AA \, (squares),
           B = 3 \AA \, (discs), B = 4 \AA \, (triangles) and
           B = 5 \AA \, (diamonds). Inversion in the sign of the
           slopes of tangential lines is seen close to the boiling point. }
\end{figure}

Equally interesting than unit volume entropies are the changes observed 
for full size cavitation, that is trends due to the creation of the entire
cavity volume. Figure \ref{fig2} shows the temperature trend of cavitation 
free energies of small sized hydrophobic solutes without scaling to the
unit volume. Tangential lines formed at room temperature data points again 
exhibit positive slopes thus again indicating a drop in entropy.  However, 
the magnitude of the slope becomes larger with increasing cavity size (e.g. 
compare slopes of tangential lines for small-sized cavities, squares 
and discs in Figure \ref{fig2}, to tangential slopes of larger-sized 
cavities, triangles and diamonds in Figure \ref{fig2}). 
Inversion in the sign of the slopes of tangential lines is seen close to 
the boiling point. Quantification of individual enthalpic and entropic 
contributions is given in Table \ref{table2}. Small negatively signed 
values of $\Delta$H are likely to result from numerical processing and 
should rather not taken to represent physical meaning. As may be seen from 
Table \ref{table2}, the process is largely dominated from entropy and only 
to a minor extent enthalpic. Therefore we need to conclude that the change 
in entropy is steadily decreasing the larger the hydrophobic solute becomes.

\subsection{ Elevated temperature simulations of molecular liquid water
             largely resemble the experimentally observed temperature     
             dependence of the liquid water density }
\begin{figure}[ht]
\centering
\includegraphics[scale=1.6]{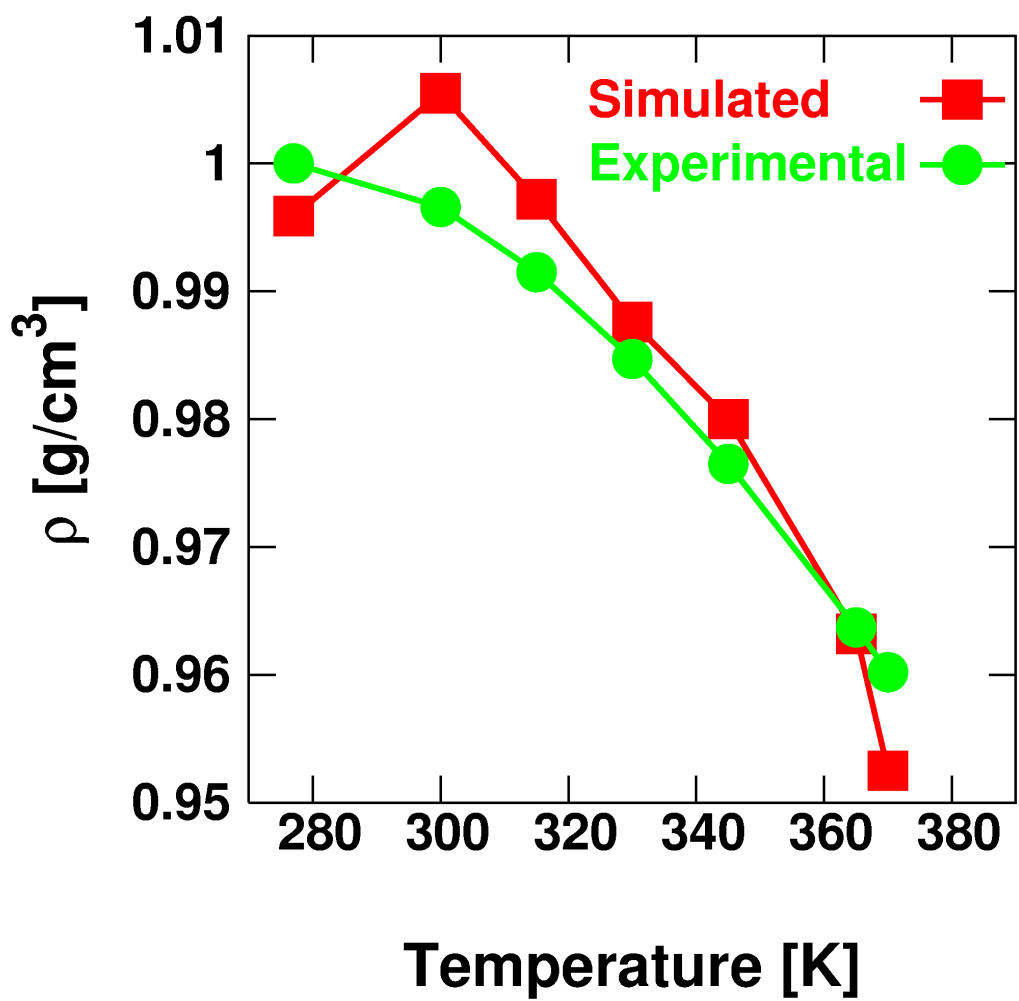}
\caption[  Comparison of simulated and measured macroscopic
           density, $\rho$, of liquid water at various temperatures. The
           simulated water densities (squares) are derived from
           $\lambda$=0.5 MD/FEP calculations that take place
           quasi-unperturbed \cite{hoefi1}. The experimental densities
           (discs) are obtained from standard tabulations
           \cite{thermexcel}. Greatest deviations from experimental
           values occur at 300 K and 370 K. The root mean square deviation
           is 0.0056 g\,cm$^{-3}$. ]
           { \label{fig3} Comparison of simulated and measured macroscopic
           density, $\rho$, of liquid water at various temperatures. The
           simulated water densities (squares) are derived from
           $\lambda$=0.5 MD/FEP calculations that take place
           quasi-unperturbed \cite{hoefi1}. The experimental densities
           (discs) are obtained from standard tabulations
           \cite{thermexcel}. Greatest deviations from experimental
           values occur at 300 K and 370 K. The root mean square deviation
           is 0.0056 g\,cm$^{-3}$. }
\end{figure}

In order to assure proper model description of intermolecular forces at  
temperatures other than room temperature, the macroscopic liquid water 
density is extracted from each simulation performed at a certain target 
temperature. Figure \ref{fig3} shows a comparison between simulation
data (squares) and experimental measurements \cite{thermexcel} (discs). 
Largest deviations occur at 300 K and 370 K with a maximum unsigned 
error of 0.01 $\frac{\rm g}{\rm cm^{3}}$ appearing at 300 K. The root mean 
square deviation amounts to 0.0056 $\frac{\rm g}{\rm cm^{3}}$. Given the 
rather close match at temperatures close to the boiling point, the overall 
rating of the simulations concerning intermolecular interactions at elevated 
temperatures must be considered very satisfactory. Consequently, taking the 
present data to form T-derivatives in order to determine entropic changes 
seems to be a valid approximation.

\subsection{ Rather close quantitative agreement between LCW predictions
             of the unit volume entropic change for the solvation of 
             small-sized hydrophobic spheres and results derived from present 
             computer simulations }
Graphical extrapolation of unit volume $\Delta$S values from the slopes of 
tangents formed at room temperature data points in temperature dependence 
plots of the unit volume $\Delta$G$^{cav}$ (Figure \ref{fig1}) yields the 
following values: 
$\Delta$S$_{\rm B=2\AA} \approx$  
-0.00027 
$\frac{\rm kcal/mol}{\rm {\rm \AA}^{3} K}$,
$\Delta$S$_{\rm B=3\AA} \approx$  
-0.00020
$\frac{\rm kcal/mol}{\rm {\rm \AA}^{3} K}$,
$\Delta$S$_{\rm B=4\AA} \approx$
-0.00015
$\frac{\rm kcal/mol}{\rm {\rm \AA}^{3} K}$ 
and
$\Delta$S$_{\rm B=5\AA} \approx$
-0.00012
$\frac{\rm kcal/mol}{\rm {\rm \AA}^{3} K}$.
A similar evaluation of the data presented in \cite{huang}
leads to comparable values, i.e.
$\Delta$S$_{\rm B=3\AA}^{\rm Huang,Chandler} \approx$
-0.00007
$\frac{\rm kcal/mol}{\rm {\rm \AA}^{3} K}$.
Thus present computer simulation data not only show qualitative agreement
with LCW theory on the hydrophobic effect of small-sized solutes, but also
lead to comparable results in absolute numbers of unit volume entropic 
changes.

\subsection{ Present computer simulations are limited to the small-size
             domain of hydrophobic hydration }
\begin{figure}[ht]
\centering
\includegraphics[scale=1.6]{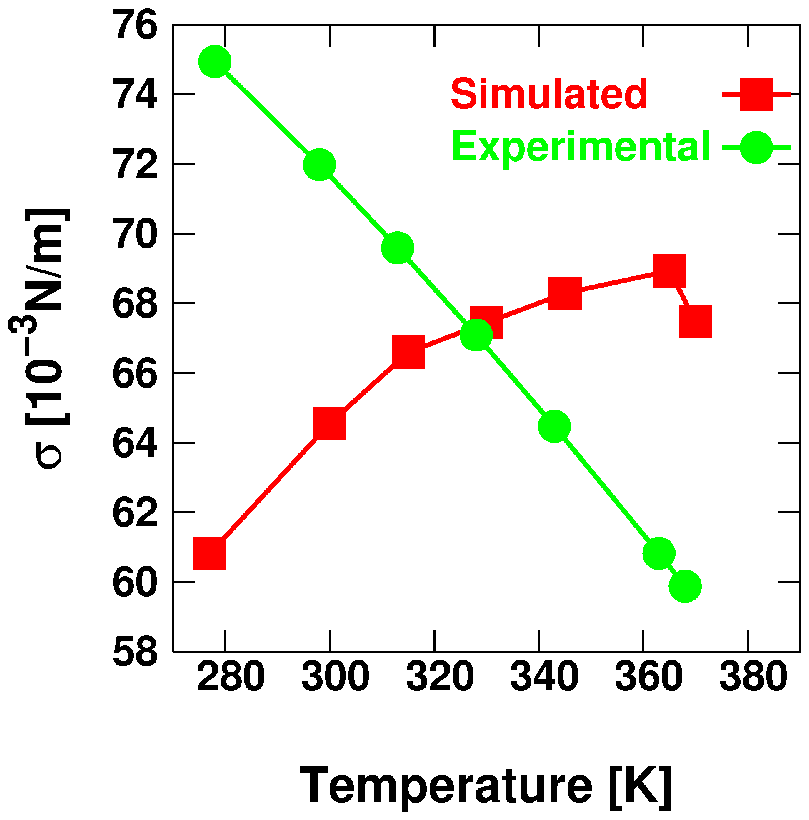}
\caption[  Comparison of simulated and measured surface
           tension, $\sigma$, of liquid water at various temperatures.
           The calculated values of $\sigma$ (squares) are due to
           $\lim\limits_{{\rm B} \to \infty}\Delta$ 
           G$^{cav}/\textcolor{red}{(}4\pi$ B$^2$\textcolor{red}{)} with 
           B=100 \AA. Experimental reference data (discs) have been 
           obtained from standard tabulations
           \cite{iapws}. Although qualitatively comparable, the
           calculated values of $\sigma$ exhibit inverse temperature
           profiles, which might be an artefact stemming from
           large-scale extrapolation. ]
           { \label{fig4} Comparison of simulated and measured surface
           tension, $\sigma$, of liquid water at various temperatures.
           The calculated values of $\sigma$ (squares) are due to
           $\lim\limits_{{\rm B} \to \infty}\Delta$ 
           G$^{cav}/\textcolor{red}{(}4\pi$ B$^2$\textcolor{red}{)} with 
           B=100 \AA. Experimental reference data
           (discs) have been obtained from standard tabulations
           \cite{iapws}. Although qualitatively comparable, the
           calculated values of $\sigma$ exhibit inverse temperature
           profiles, which might be an artefact stemming from
           large-scale extrapolation. }
\end{figure}

rPA coefficients obtained from MD/FEP calculations in AMOEBA water
are strictly valid only in the domain of existing FEP data. It is reasonable 
to extrapolate into the extended cavity size domain due to the smoothness of 
the data. However, after about B=5 \AA\, (twice the radius of the largest 
accumulated perturbation) any further usage is certainly speculative. 
Therefore it was surprising to see indications of general applicability 
even in largely extended cavity domains \cite{hoefi}. In order to probe the 
quality of large scale extrapolation the temperature trend of the surface 
tension, $\sigma$, is shown in Figure \ref{fig4} and compared to
experimental data \cite{iapws}. Approximation of $\sigma$ from the present 
rPA data is critical (see Discussion). It involves limit value consideration 
of $\lim\limits_{{\rm B} \to \infty}\Delta$ G$^{cav}/\textcolor{red}{(}4\pi$ 
B$^2$\textcolor{red}{)} \cite{hoefi} 
and the presented data is due to setting B=100 \AA, hence clearly an 
extrapolation to large scale. Figure \ref{fig4} shows that the calculated 
surface tensions are close to the experimental values but do not reproduce 
the correct trend with increasing temperature, demonstrating the anticipated 
uncertainty with large-size approximations.

\subsection{ \textcolor{green}{ Analysis of the hydrogen bond network as 
              determined by the ensemble of snapshot structures obtained
              from AMOEBA based MD simulations goes hand in hand with 
              recently reported findings of experimental as well as 
              theoretical studies                               }         }
\textcolor{green}{ MD/FEP runs carried out at $\lambda=0.5$ 
(quasi-unperturbed) are examined with respect to structural relationships 
maintained between individual water molecules. Corresponding sets of saved 
snapshot structures of individual water boxes at T=300 K and T=370 K are
analyzed. The three basic {\it Radial Distribution Functions} (RDF) that
characterize the ``fine structure'' of bulk water are derived and 
corresponding trends shown in the Supplementary Material, Figures A-C 
(300 K) and Figures D-F (370 K). RDFs from the present data reproduce those 
reported in the original AMOEBA paper (see \cite{ren} Figures 5 -- 7) very 
well. The RDF of the oxygen-oxygen distance, g$_{\rm OO}$(r), shows the 
characteristic immediate neighborhood peak at 2.9 \AA~and a second very 
diffuse peak around 4.6 \AA~which is hardly distinguishable from the 
baseline. The second peak in the g$_{\rm OH}$(r) --- due to pairs of water 
molecules directly associated via hydrogen bonds (HBs) --- appears at 1.9 
\AA~in the present data. The g$_{\rm HH}$(r) exhibits a characteristic 
second peak at 2.5 \AA. Following the procedure devised in \cite{xenides} 
for geometrical constraints defining a particular HB we determine an average 
number of established HBs of 1.6 per water molecule in the 300 K simulation 
data. The frequency of tetrahedral coordination mediated by HBs is only on 
the order of 2 \% at this simulation temperature. This seems to be 
consistent with the experimental study of Wernet et al \cite{wernet} where
the conclusion was drawn that bulk water predominantly exists in a state 
with 2 HBs established per water molecule. The present data is also in close 
agreement with the MP2 results of the QM/MM study by Xenides et al 
\cite{xenides}. Significant changes observed at the high temperature 
simulation of T=370 K include the entire loss of the second broad shoulder 
around  4.6 \AA~ in the g$_{\rm OO}$(r), the decrease of the average number 
of HBs to 1.4 per water molecule and the general loss of tetrahedral 
coordinations mediated by HBs.                                                }

\section{ Discussion }
The present study has employed the high level polarizable AMOEBA force
field for computer simulation of liquid water \cite{ren} at several 
temperatures and normal pressure. We are interested in the free energy cost 
of creation of small-sized cavities, which are model constructs for 
hydrophobic molecules. A first requirement was to show that intermolecular 
relationships are still maintained at reasonable physical conditions upon
temperature increase \cite{paschek}. A rather direct evidence in this regard 
is the relatively close match between simulated and experimental data of the 
liquid water density shown in Figure \ref{fig3}. Next the temperature profile 
of $\Delta$G was used to get estimates for the unit volume $\Delta$S (Figure 
\ref{fig1}) as well as for the full size $\Delta$S of cavitation (Figure 
\ref{fig2}). Computer simulation values for $\Delta$S can be derived from 
the negative slope of the tangents constructed at a certain temperature. In 
so doing the $\Delta$S for room temperature hydrophobic hydration of 
small-sized molecules was shown to exhibit a negative sign, which was 
predicted previously from LCW-theory \cite{lum}. The absolute value of 
the change in cavitation entropy, $\Delta$S, becomes larger with increasing 
size of the hydrophobic molecule (see Figure \ref{fig2} and Table 
\ref{table2}). Tendency inversion is observed only at elevated temperatures 
close to the boiling point.

In general, the LCW picture of the unit volume hydrophobic effect of 
small-sized solutes is very well reproduced (see for example Figure 2 in 
\cite{huang}). Small differences include, the exact position of the 
maximum, which occurs closer to the boiling point in the present data,     
the quantification of the unit volume $\Delta$S, which leads to higher 
values in the present study and the continuity of the onset of the curves 
at temperatures close to the freezing point, which is steady here, but shows 
some inversion at B=4 \AA \, solutes in LCW-theory. Extrapolation of the 
present data to very large solute sizes is impossible as seen from the 
wrong prediction of the temperature profile of the surface tension 
(Figure \ref{fig4}). The present approach has made use of a very expensive 
computational technique (Ewald sum MD/FEP within the OST) and could only 
be carried out under massive employment of grid computing \cite{AutGrid}. 
All final rPA coefficients corresponding to all the considered temperatures 
are summarized in Table \ref{table1}. It is interesting to note, that all
present $k_0$, $k_1$, $k_2$ coefficients derived for 300 K lead to only 
minor alterations in the data presented recently in \cite{hoefi} that were 
based on non-Ewald simulations, but application of plain periodic boundary 
conditions instead \cite{hoefi1}.

An interesting observation made in the present work is the fact that the 
overall entropy of hydrophobic hydration is steadily decreasing with growing 
solute size (see Figure \ref{fig2} and Table \ref{table2}). This is 
somewhat contradictory with the LCW-picture of the anticipated change in 
the sign of $\Delta$S when the solutes are thought to cross over from small 
length scales (volume dependence) to large length scales (surface area 
dependence). The critical crossover dimension was approximated to be on the 
order of 10 \AA \, of radial extension of the hydrophobic solute \cite{lum}. 
On the other hand, strictly speaking, the crossover region is beyond the 
reach of the current data set and firm statements can only be made upon 
extension of the present approach into the critical domain of crossover 
length scales. This would however involve an even bigger initiative of 
supercomputing with unforeseeable complexity. It is still interesting to 
note that an extension of the rPA approach into medium length scales 
was recently shown to be not entirely unreasonable when operating with the 
solvent excluded volume \cite{hoefi}.

Predicting the temperature dependence of the surface tension (see Figure 
\ref{fig4}) from present rPA data failed. First of all it should be noted, 
that surface tension per se is appropriate for the description of water 
embedded macroscopic bubbles. These bubbles are fundamentally different 
to cavities of comparable size, because the latter are strictly empty. 
Therefore, all what makes a bubble stabilize, i.e. the vapour-like molecules 
in the interior bouncing back and fourth against the bubble walls, will be 
completely missing in a cavity, because the cavity interior is void. 
Nevertheless, the equivalence in units of surface-normalized cavitation 
free energies on the one side and surface tensions on the other side, 
makes it attractive to try the prediction of the latter via 
$\lim\limits_{{\rm B} \to \infty}\Delta$ G$^{cav}/\textcolor{red}{(}4\pi$ 
B$^2$\textcolor{red}{)}. However,
the energy in the context of surface tension implies an energy required 
to maintain the liquid-vapour coexistence of a bubble, while it expresses 
an amount of clearance work to create the empty space in a cavity. These
two different types of energies should not be confused with each other and 
in principle, need not even be correlated. Thus the rather successful
match of $\lim\limits_{{\rm B} \to \infty}\Delta$ 
G$^{cav}/\textcolor{red}{(}4\pi$ B$^2$\textcolor{red}{)} with the experimental 
surface tension of water reported in \cite{hoefi} could also be purely 
fortuitous. On the other hand it could also explain why surface tension 
based solvation models \cite{cramer} and LCW-theory have become so 
successful.

LCW theory --- like the earlier concepts introduced by Kauzmann 
\cite{kauzmann} and Tanford \cite{tanford} --- build upon oil/water surface 
tension and explain the switch in sign of $\Delta S$ for the hydration of 
hydrophobic solutes of increasing size by the experimental temperature 
dependence of surface tensions. LCW in particular has proclaimed that such 
a switch in sign would occur at spherical volumes of radial dimension on 
the order of 10 \AA. Taking up this idea, $\Delta S$ values in the vicinity 
of the 10 \AA \, domain should smoothly become smaller and smaller before they 
reach zero and then start to assume negative values. A plot of $\Delta S$ in
this critical region should therefore certainly not exhibit characteristics 
of a monotonic growth, otherwise the transition would have to occur 
discontinuously. In the present study, although still operating far away 
from the critical domain of 10 \AA, we do however observe signatures of a
monotonic growth (see Table \ref{table2}). This will be of particular 
interest to examine in closer detail from extended studies of the type 
presented here.

\section{ Conclusion }
In summary, evidence has been presented that the hydrophobic effect of
small-sized solutes is due to the drop in entropy. The insight gained here 
is also an encouraging sign that high level force fields are approaching 
now standards that allow one to study effects beyond the reach of 
experimental methods.

{\bf\large Acknowledgements }
The authors are grateful for having been granted all necessary computer 
time by the Austrian Grid \cite{AutGrid}. This work was supported in part 
by the National Institutes of Health Grant GM62838. The authors thank 
Petra Bareis and Roland Felnhofer from the Novartis Institutes for 
BioMedical Research, Vienna, for their collaborative spirit.  
\pagebreak


\clearpage
\begin{table}
\begin{center}
\caption[]{\label{table1}Summary of the data resulting from free energy 
           perturbation calculations of repulsive spheres in water at 
           various simulation temperatures. Equilibration values of liquid 
           water densities (second column) are derived from 
           $\lambda$ = 0.5 runs (quasi-unperturbed) after neglecting an 
           initial fraction of the data set (fourth column). The data are 
           fitted with polynomials of 2$^{\rm nd}$ degree \cite{hoefi2} 
           and resulting rPA-coefficients $k_0$, $k_1$ and $k_2$ are 
           included in columns 5-7. }
\vspace{0.5cm}
\begin{tabular}{lcccccc}
\\
\hline\hline
       \\
       T                                                  &
       $\rho_{\rm sim}$                                   &  
       $\rho_{\rm exp}$                                   & 
       Initial                                            & 
       \multicolumn{3}{c}{rPA-Coefficients \cite{hoefi}}  \\ 
                                                          &
                                                          &    
                                                          &    
       Discard                                            &
       $k_0$                                              &                    
       $k_1$                                              &                    
       $k_2$                                              \\                    
       $[$K$]$                                            &
       $[\frac{\rm g}{\rm cm^{3}}]$                       &
       $[\frac{\rm g}{\rm cm^{3}}]$                       &
       $[$ps$]$                                           &
       $[{\rm kcal/mol}]$                                 & 
       $[\frac{\rm kcal/mol}{\textrm{ \AA }}]$            &
       $[\frac{\rm kcal/mol}{{\textrm{ \AA }}^2}]$        \\
       \\
\hline
\\
277  &  0.995  &  1.000  &  40  &  
0.456 $\pm$ 0.015  &  -1.602 $\pm$ 0.062  &  1.116 $\pm$ 0.082  \\
300  &  1.006  &  0.996  &  40  &  
0.427 $\pm$ 0.012  &  -1.594 $\pm$ 0.044  &  1.183 $\pm$ 0.046  \\
315  &  0.997  &  0.992  &  40  &  
0.418 $\pm$ 0.041  &  -1.613 $\pm$ 0.060  &  1.220 $\pm$ 0.022  \\
330  &  0.988  &  0.985  &  40  &  
0.398 $\pm$ 0.008  &  -1.607 $\pm$ 0.045  &  1.236 $\pm$ 0.052  \\
345  &  0.980  &  0.977  &  40  &  
0.405 $\pm$ 0.011  &  -1.607 $\pm$ 0.045  &  1.251 $\pm$ 0.045  \\
365  &  0.963  &  0.964  &  40  &  
0.383 $\pm$ 0.015  &  -1.580 $\pm$ 0.026  &  1.262 $\pm$ 0.038  \\
370  &  0.953  &  0.960  &  50  &  
0.411 $\pm$ 0.009  &  -1.611 $\pm$ 0.037  &  1.236 $\pm$ 0.029  \\
\\ \\
\hline\hline
\end{tabular}
\end{center}
\end{table}

\clearpage
\begin{table}
\begin{center}
\caption[]{\label{table2}Decomposition of cavitation free energies 
           $\Delta$G$^{cav}$ (column to the right) of small-sized spherical 
           volumes of radius $B$ (leftmost column) into entropic (third 
           column) and enthalpic (second column) fractions. The 
           defragmentation is based on initial finite difference 
           determination of $\Delta$S from the temperature trend of 
           $\Delta$G$^{cav}$ (see Figure \ref{fig2}) with subsequent 
           trivial derivation of $\Delta$H. The analysis is made for room 
           temperature behaviour (T=300 K).}
\vspace{0.5cm}
\begin{tabular}{lccc}
\\
\hline\hline
       \\
       B                             &
       $\Delta$H                     &  
       -T$\Delta$S                   & 
       $\Delta$G$^{cav}$             \\ 
       $[$\AA$]$                     &
       $[{\rm kcal/mol}]$            &
       $[{\rm kcal/mol}]$            &
       $[{\rm kcal/mol}]$            \\
       \\
\hline
\\
2    &  -0.728  &  2.699  &  1.972   \\
3    &  -0.335  &  6.628  &  6.293   \\
4    &   0.798  & 12.183  & 12.981   \\
5    &   2.672  & 19.364  & 22.036   \\
\\ \\
\hline\hline
\end{tabular}
\end{center}
\end{table}

\clearpage
\listoffigures

\clearpage

\clearpage

\clearpage

\clearpage


\clearpage

\begin{thebibliography}{99}
\bibitem[1]{chandler}
D. Chandler, {\it Nature}, 2005, {\bf 437}, 640.
%
\bibitem[2]{liu}
P. Liu, X. Huang,  R. Zhou, B. J. Berne, 
{\it Nature}, 2005, {\bf 437}, 159.
%
\bibitem[3]{despa}
F. Despa, A. Fern\a'{a}ndez, R. S. Berry, 
{\it Phys. Rev. Lett.},  2004, {\bf 93}, 228104.
%
\bibitem[4]{lum}
K. Lum, D. Chandler, J. D. Weeks, 
{\it J. Phys. Chem. B}, 1999, {\bf 103}, 4570.
%
\bibitem[5]{pratt}
L. R. Pratt, A. Pohorille, 
{\it Proc. Natl. Acad. Sci.} USA, 1992, {\bf 89}, 2995.
%
\bibitem[6]{hummer}
G. Hummer,  S. Garde, A. E. Garcia, A. Pohorille, L. R. Pratt, 
{\it Proc. Natl. Acad. Sci.} USA, 1996, {\bf 93}, 8951.
%
\bibitem[7]{floris}
F. M. Floris, M. Selmi, A. Tani, J. Tomasi, 
{\it J. Chem. Phys.},  1997, {\bf 107}, 6353.
%
\bibitem[8]{choudhury}
N. Choudhury, B. M. Pettitt, 
{\it J. Am. Chem. Soc.},  2005, {\bf 127}, 3556.
%
\bibitem[9]{ashbaugh}
H. S. Ashbaugh, M. E. Paulaitis, 
{\it J. Am. Chem. Soc.}, 2001, {\bf 123}, 10721.
%
\bibitem[10]{zwanzig}
R. W. Zwanzig, 
{\it J. Chem. Phys.}, 1954, {\bf 22}, 1420.
%
\bibitem[11]{postma}
P. M. Postma, H. J. C. Berendsen, J. R. Haak, 
{\it Farad. Symp. Chem. Soc.}, 1982, {\bf 17}, 55.
%
\bibitem[12]{simonson}
T. Simonson,  G. Archontis,  M. Karplus, 
{\it Acc. Chem. Res.}, 2002, {\bf 35}, 430.
%
\bibitem[13]{tomasi}
J. Tomasi, B. Mennucci, R. Cammi, 
{\it Chem. Rev.}, 2005, {\bf 105}, 2999.
%
\bibitem[14]{hoefi}
S. H\"ofinger, F. Zerbetto, 
{\it Chem. Soc. Rev.}, 2005, {\bf 34}, 1012.
%
\bibitem[15]{paschek}
D. Paschek, 
{\it J. Chem. Phys.}, 2004, {\bf 120}, 6674.
%
\bibitem[16]{paschek1}
P. E. Krouskop, J. D. Madura, D. Paschek, A. Krukau, 
{\it J. Chem. Phys.}, 2006, {\bf 124}, 016102-[1,2].
%
\bibitem[17]{zhou}
R. Zhou, X. Huang, C. J. Margulis, B. J. Berne, 
{\it Science}, 2004, {\bf 305}, 1605.
%
\bibitem[18]{pande}
M. R. Shirts, V. S. Pande, 
{\it J. Chem. Phys.}, 2005, {\bf 122}, 134508.
%
\bibitem[19]{jorgensen}
W. L. Jorgensen, J. Chandrasekhar, J. D. Madura, R. W. Impey,  
M. L. Klein, 
{\it J. Chem. Phys.}, 1983, {\bf 79}, 926.
%
\bibitem[20]{berendsen}
H. J. C. Berendsen, J. P. M. Postma, W. F. van Gunsteren,  
J. Hermans,  
{\it Intermolecular Forces;} Reidel Publishing Company: Dordrecht, 
the Netherlands, 1981.
%
\bibitem[21]{ren}
P. Ren, J. W. Ponder, 
{\it J. Phys. Chem. B}, 2003, {\bf 107}, 5933.
%
\bibitem[22]{ren1}
P. Ren, J. W. Ponder, 
{\it J. Phys. Chem. B}, 2004, {\bf 108}, 13427.
%
\bibitem[23]{hoefi1}
S. H\"ofinger, F. Zerbetto, 
{\it Chem. Eur. J.}, 2003, {\bf 9}, 566.
%
\bibitem[24]{ghosh}
T. Ghosh, A. E. Garcia, S. Garde, 
{\it J. Chem. Phys.}, 2002, {\bf 116}, 2480.
%
\bibitem[25]{rajamani}
S. Rajamani, T. M. Truskett, S. Garde, 
{\it Proc. Natl. Acad. Sci.} USA, 2005, {\bf 102}, 9475.
%
\bibitem[26]{benjamin}
\textcolor{red}{B. Almeida, R. Mahajan, D. Kranzlm\"uller, J. Volkert, 
S. H\"ofinger,
{\it Lect. Notes. Comp. Sci.}  2005, {\bf 3666}, 433. }
%
\bibitem[27]{AutGrid}
J. Volkert, D. Kranzlm\"uller,
{\it OCG Journal}, 2005, {\bf 1}, 4.
%
\bibitem[28]{hoefi2}
S. H\"ofinger, F. Zerbetto, 
{\it Theor. Chem. Acc.}, 2004, {\bf 112}, 240.
%
\bibitem[29]{huang}
D. M. Huang, D. Chandler, 
{\it Proc. Natl. Acad. Sci.} USA, 2000, {\bf 97}, 8324.
%
\bibitem[30]{thermexcel}
http://www.thermexcel.com
ThermExcel
Copyright \copyright \, 2003-2004 by ThermExcel
%
\bibitem[31]{iapws}
http://www.iapws.org/relguide/surf.pdf 
{\it IAPWS, International Association for the Properties of Water 
and Steam.} 1994
%
\bibitem[32]{cramer}
C. J. Cramer, D. G. Truhlar, 
{\it Chem. Rev.}, 1999, {\bf 99}, 2161.
%
\bibitem[33]{ponder}
J. W. Ponder, 
TINKER 
Copyright \copyright \, 1990-2004 by Jay William Ponder
%
\bibitem[34]{berendsen1}
H. J. C. Berendsen, J. P. M. Postma, W. F. van Gunsteren, A. DiNola,
J. R. Haak, 
{\it J. Chem. Phys.}, 1984, {\bf 81}, 3684.
%
\bibitem[35]{andersen}
\textcolor{red}{H. C. Andersen, 
{\it J. Chem. Phys.}, 1980, {\bf 72}, 2384. }      
%
\bibitem[36]{cheatham}
\textcolor{red}{S. C. Harvey, R. K. Z. Tan, T. E. Cheatham III,              
{\it J. Comp. Chem.}, 1998, {\bf 19}, 726. }
%
\bibitem[37]{beutler}
T. C. Beutler, A. E. Mark, R. C. van Schaik, P. R. Gerber,   
W. F. van Gunsteren, 
{\it Chem. Phys. Lett.}, 1994, {\bf 222}, 529. 
%
\bibitem[38]{simonson1}
T. Simonson, 
{\it Mol. Phys.}, 1993, {\bf 80}, 441. 
%
\bibitem[39]{kauzmann}
W. Kauzmann, 
{\it Adv. Protein Chem.}, 1959, {\bf 14}, 1.
%
\bibitem[40]{tanford}
C. Tanford, 
{\it The hydrophobic effect --- formation of micelles and biological 
membranes;} Wiley: New York, 1973.
%
\bibitem[41]{wernet}
\textcolor{red}{ Ph. Wernet, D. Nordlund, U. Bergmann, M. Cavalleri, 
M. Odelius, H. Ogasawara, L.\AA. N\"aslund, T.K. Hirsch, L. Ojam\"ae, 
P. Glatzel, L.G.M. Pettersson, A. Nilsson,
{\it Science}, 2004, {\bf 304}, 995. }
%
\bibitem[42]{xenides}
\textcolor{red}{ D. Xenides, B.R. Randolf, B.M. Rode,
{\it J. Chem. Phys.}, 2005, {\bf 122}, 174506. }
%
\end{thebibliography}
\end{document}